# Bifacial thermophotovoltaic energy conversion

A. Datas[*]

Instituto de Energía Solar, Universidad Politécnica de Madrid, Avenida Complutense, 30, 28040, Madrid, Spain

* Corresponding author: a.datas@upm.es

**Abstract**

Thermophotovoltaic (TPV) energy conversion efficiency has recently surpassed 30%. The key behind such high efficiency is the inclusion of a highly efficient mirror in the rear of the TPV cell that turns back to the thermal emitter the outband energy photons. Efficiencies over 50% could be theoretically attainable by approaching a mirror reflectance of 100%. However, the very few percent of outband absorption significantly deteriorate the conversion efficiency. Thus, current research focuses on developing advance mirror designs able to reach an extreme high outband reflectance over 95%. In this article I propose a bifacial TPV cell that enables very efficient photon recycling without using mirrors and that is less sensitive to outband optical losses. The key to this design is that the cell is introduced in a thermal emitter enclosure where it is irradiated from both sides. Then, outband photons transmit through the cell and are re-absorbed in the emitter. Therefore, the optical losses linked to the mirror/cell interface are eliminated, potentially enabling higher photon recycling efficiencies. This article presents a detailed balance simulation of an edge-cooled bifacial TPV cell to demonstrate that bifacial configuration enables higher conversion efficiencies and twice much as power density than monofacial designs, the latter being a remarkable advantage for moderate temperature and low-cost TPV power generation. Therefore, bifacial TPV cells are appealing for developing practical high-efficient and low-cost TPV devices for power generation in an extended range of heat source temperatures.

**Introduction**

Thermophotovoltaics (TPV) is the direct conversion of thermal radiation into electricity through the photovoltaic effect [1], [2]. In a TPV device (see panel a in Figure 1), photons are radiated by a thermal emitter and directed towards a closely spaced TPV cell, which is typically made of a low bandgap semiconductor. Within the cell, photons with wavelengths shorter that of the semiconductor bandgap ($\lambda < \lambda_g$ in panels a and d of Figure 1) are absorbed and produce electron-hole pairs. Then, electrons and holes are collected in separate contacts (an electron transport layer – ETL – and a hole transport layer – HTL –), subsequently producing an external current. In contrast to conventional solar cells, a high TPV conversion efficiency can be achieved by turning back to the heat source the low energetic photons (i.e., those with $\lambda > \lambda_g$ in panels a and d of Figure 1) that do not contribute to the electron-hole pair generation. In this way, some of these photons are recycled and do not contribute to the overall heat loss. As a matter of fact, in the recent years, the development of highly efficient mirrors in the back side of TPV cells, the so-called back-surface reflectors (BSR) (panel *a* in Figure 1), have led to record TPV cell conversion efficiencies in the range of 30 to 40 % [3]–[6]. Further improvements in these BSR could lead to efficiencies over 50 %, especially if combined with multijunction cell architectures and extreme high temperature emitters [5], [6].

Despite BSR-TPV cell architectures have been proven effective to boost the TPV cell efficiency, current designs are facing unavoidable sub-bandgap (outband) photon absorption, which mostly occurs in the metallic reflector due to the waveguide modes that exist at the semiconductor-metal interface [4], [5], [7]. Therefore, sophisticated and somehow challenging reflector designs, e.g. air-bridge configurations [3], are needed to reach outband reflectance over 95% and achieve the highest conversion efficiencies. Moreover, most of the TPV devices reported to date produce relatively low power densities (e.g., < 1 W/cm$^2$ at emitter temperatures < 1200ºC [8]) which is attributed to the small fraction of in-band spectral irradiance ($\lambda < \lambda_g$ in panel d of Figure 1) absorbed in the TPV cell. Increasing the power density is essential to reduce the cost of TPV power and make it profitable in a wide range of heat source temperatures. Proposed solutions to increase the power density include near-field arrangements [9], light-pipes [10], hybrid thermionic-photovoltaics [11], or electroluminescent heat pumps (thermophotonics) [12], but none of them has been yet found feasible to boost the TPV power density at a relevant scale.

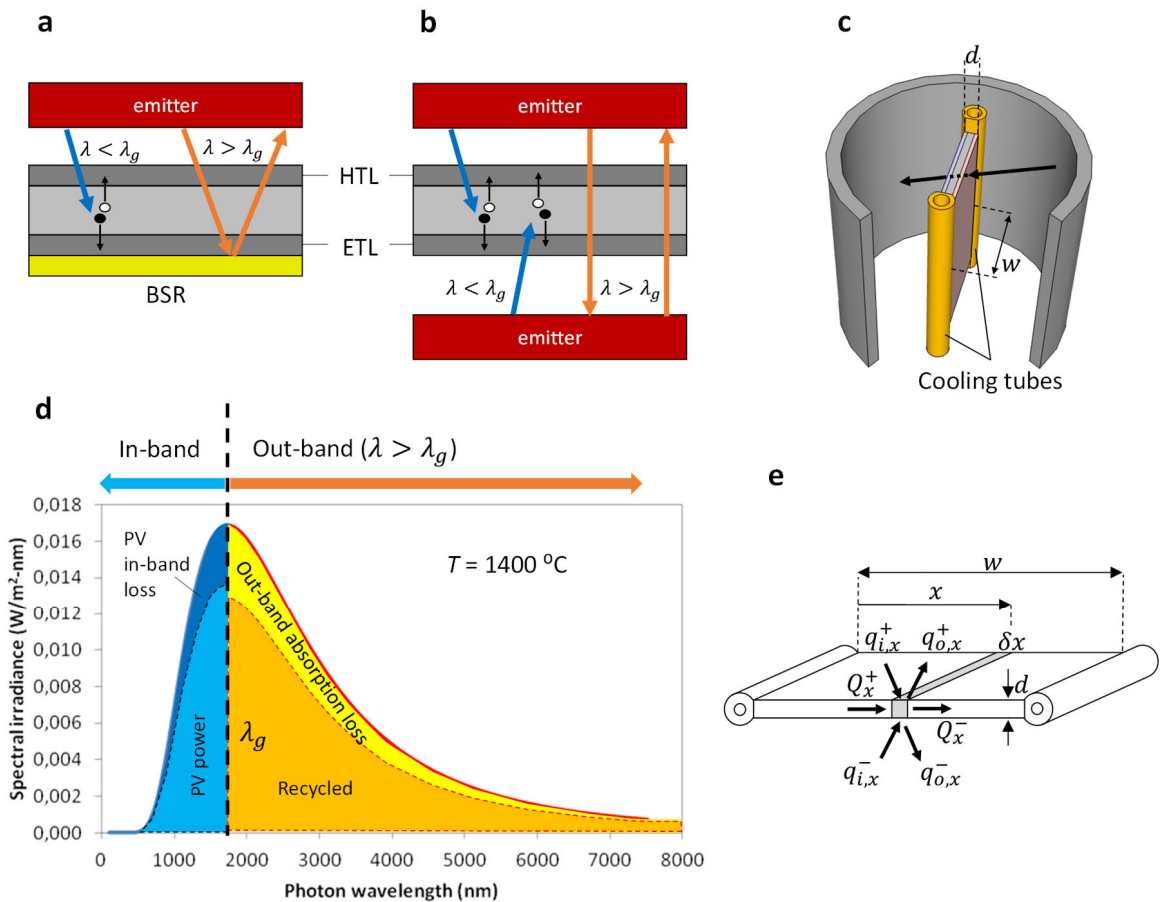

Figure 1 Panels *a* and *b* show the two kinds of TPV cells analyzed in this work - BSR (panel a) and bifacial (panel b) – having both electron- and hole-selective transport layers (ETL and HTL); Panel c shows the edge-cooled bifacial TPV cell within a thermal emitter enclosure, which is the specific bifacial design analyzed in this article. Panel d shows the spectral irradiance of the blackbody emitter indicating the wavelength of the semiconductor bandgap ($\lambda_g$) that delimits in-band and out-band radiation and the portions of the radiation that are lost, recycled, or converted into electricity. Panel e shows the heat fluxes in a differential element of the bifacial TPV cell, including incoming / outgoing conductive heat fluxes ($Q_x^+$ and $Q_x^-$) and incoming / outgoing radiative heat fluxes in both sides of the cell ($q_{i,x}^+, q_{i,x}^-, q_{o,x}^+, q_{o,x}^-$).

In this article a bifacial TPV cell design that enables higher conversion efficiency and provides around twice as much power density than conventional BSR-TPV cells is proposed. The bifacial TPV cell (see panel *b* in Figure 1) is designed to be introduced in a thermal emitter enclosure to collect thermal radiation from both sides. In this arrangement, outband photons ($\lambda > \lambda_g$) are transmitted through the cell and reabsorbed in the emitter. Therefore, outband photon recycling is accomplished without using mirrors, and the optical losses linked to the mirror/cell interface are eliminated, potentially enabling higher overall photon recycling efficiencies. Moreover, as the cell is illuminated from both sides, the cell is able to produce twice the power than conventional BSR-TPV cell designs. Other practical benefit is that the cells can be easily interconnected in highly packed modules with a minimum area dedicated to the interconnection; thus, potentially enabling very low cell-to-module efficiency losses.

Because both sides of the bifacial cell will be exposed to a high irradiance, a main challenge of bifacial TPV concept is the cell cooling. To that end, this article assesses a possible design where the edges of the bifacial cell are attached to highly reflective tubes through which a cooling fluid is made to flow (panels c and e in Figure 1). It will be seen that this arrangement enables an effective cooling, provided that the cell has a minimum thickness and a maximum width, whose values mostly depend on the emitter temperature and the outband optical absorption in the cell.

This article assesses the fundamentals of bifacial TPV energy conversion, focusing on the impact of the two main fundamental kinds of losses, i.e., outband absorption and non-radiative recombination. The output performance (efficiency and power density) of bifacial TPV converters will be evaluated and compared to conventional BSR-TPV cell designs. We will see that bifacial TPV cells can reach higher conversion efficiencies even with higher outband optical losses and internal non-radiative recombination. Remarkably, bifacial TPV cells are less sensitive to both kinds of losses and can produce approximately twice the power density than conventional BSR-TPV devices. Therefore, bifacial TPV cells are appealing for high-efficient and low-cost TPV power generation in an extended range of heat source temperatures.

**Theoretical background and methodology**

The edge-cooled bifacial TPV cell described above is modelled in this work using the detailed balance theory. Proposed by Shockley and Queisser in 1961 [13], this method assumes infinite mobility for the carriers. Thus, it is frequently used to calculate upper bounds for the conversion efficiency of novel PV concepts, including TPV [14]. The heat flux and temperature gradient in the cell are calculated by establishing the heat energy balance at a differential element d$x$ (see panel $e$ of Figure 1) as

$$k_{th} d \frac{d^2 T}{dx^2} = -Q(x) \qquad (1)$$

being $k_{th}$ the cell thermal conductivity, $d$ the cell thickness, and $Q(x)$ the net thermal power density generated, in W/m$^2$, at the differential element d$x$ at a position $x$ within the cell. Assuming that the emitter is a black body and that both the cell and the emitter are Lambertian surfaces at temperatures $T_c(x)$ and $T_e$, respectively, $Q(x)$ can be expressed as

$$Q(x) = 2 \int_\lambda A(\lambda)[\dot{e}(\lambda, T_e, 0) - \dot{e}(\lambda, T_c(x), qV)] \, d\lambda - P_d \qquad (2)$$

being $A(\lambda)$ the cell' spectral absorptivity/emissivity, $P_d$ the cell's electric output power density (in W/m$^2$), and $\dot{e}(\lambda, T, \mu)$ the spectral energy flux, in W/m$^2$-m, emitted by a Lambertian black body surface with temperature $T$ and electrochemical potential $\mu$:

$$\dot{e}(\lambda, T, \mu) = \frac{2\pi h c^2}{\lambda^5} \frac{1}{\exp\left(\frac{hc/\lambda - \mu}{k_B T}\right) - 1} \tag{3}$$

where $h$ and $k_B$ are the Plank and Boltzmann constants, respectively, $c$ is the speed of light, and $\lambda$ is the photon wavelength. The cell is biased at a voltage $V$ that produces an electroluminescent radiation with electrochemical potential $\mu = qV$, being $q$ the electron charge. In this work, the bifacial cell is assumed to have a unitary absorptivity/emissivity for photons with energy higher than that of the cell's bandgap ($\varepsilon_g = hc/\lambda_g$), and a variable outband absorption ($0 < OBA_{bif} < 1$) otherwise:

$$A(\lambda) = \begin{cases} OBA_{bif}, & \lambda > \lambda_g \\ 1, & \lambda \leq \lambda_g \end{cases} \tag{4}$$

The TPV cell output power $P_d = \max(J \cdot V)$ is calculated using the detailed balance theory, following a similar approach than in [14]–[16]. According to this theory, the current-voltage characteristic of a TPV cell is given by

$$j(V) = q \int_0^{\lambda_g} \{ K_{bi}\big(\dot{n}(\lambda, T_e, 0) - \dot{n}(\lambda, T_c, qV)\big) - K_{mo}\dot{n}^2(1 - \rho_{BSR})\dot{n}(\lambda, T_c, qV) \\ - 2n^2((1 - \eta_{int})/\eta_{int})\,\dot{n}(\lambda, T_c, qV)\}d\lambda \tag{5}$$

where $K_{bi} = 2$ (1) and $K_{mo} = 0$ (1) for bifacial (monofacial BSR) TPV cells, $\dot{n}(\lambda, T, \mu) = \dot{e}(\lambda, T, \mu)/(hc/\lambda)$ is the normal photon flux emitted by a black body Lambertian surface at temperature $T$ and electrochemical potential $\mu$, $n$ is the refractive index of the semiconductor, $\rho_{BSR}$ is the reflectance of the semiconductor/BSR interface (only for monofacial BSR cells), and $\eta_{int}$ is the internal electroluminescence efficiency [16].

The TPV cell efficiency is finally obtained from the delivered output power and heat as

$$\eta_{TPV} = \frac{P_d}{P_d + Q_{tot}} \quad (6)$$

being $Q_{tot} = \frac{1}{w}\int_{x=0}^{x=w} Q(x)dx$ the total heat dissipated from the bifacial cell in W/m². For the bifacial cell, $Q_{tot}$ can be also obtained through Fourier law evaluated at the edges of the cell as $Q_{tot} = 2\frac{d}{w} \cdot k_{th} dT/dx|_{x=w}$. For monofacial BSR TPV cells $Q_{tot}$ is obtained from an energy balance in the cell, similar to equation 2, but taking also into account BSR reflectivity [14]–[16]:

$$Q_{tot,BSR} = \int_0^\infty \dot{e}(\lambda, T_e, 0)d\lambda - \int_0^{\lambda_g} \dot{e}(\lambda, T_c, qV)d\lambda - \int_{\lambda_g}^\infty \rho_{BSR}\dot{e}(\lambda, T_e, 0)d\lambda - P_d \quad (7)$$

Notice that equation 7 assumes all optical losses taking place at the mirror/reflector interface. Thus, it neglects outband absorption losses in the substrate. This is a good assumption for typical thin-film BSR TPV cell architectures [7].

The models described above can be used to calculate some key performance indicators of the bifacial and BSR TPV cells, such as the output power density, the conversion efficiency, or the average and peak cell temperature (only for bifacial cells). For the sake of simplicity, this work assumes a TPV cell made of a semiconductor with a bandgap energy of 0.66 eV (e.g., Ge or Ga$_{0.38}$In$_{0.62}$As), a constant refractive index of 3.5, and a thermal conductivity of $k_{th} = 0.6 \, W/cm$-K (typical value for Ge and InP, the latter typically used as substrate for growing Ga$_{0.38}$In$_{0.62}$As semiconductors). The thermal emitter is assumed to be a black body and the cooling system is assumed to keep a constant temperature of 27ºC at the edges of the cell. For the calculations in equation 5, bifacial cells are assumed to operate at an average temperature $T_{c,avg} = \frac{1}{w}\int_{x=0}^{x=w} T_c(x)dx$, being $w$ the cell width.

It is worth noting that the results presented in this work aim at maximizing the power density (i.e., $P_d = JV$) rather than the efficiency of the TPV cells. Maximizing the efficiency would result in higher output voltages that minimize the heat absorbed in the cells by maximizing the electroluminescent radiation turned back to the thermal emitter. However, this may lead to negligible power generation [14].

**Results and discussion**

Figure 2 shows the relative difference in the efficiency between bifacial and BSR cells as a function of the outband absorption (OBA) losses for each kind of cell (i.e., $OBA_{bif}$ for the bifacial cell and $OBA_{BSR} = 1 - \rho_{BSR}$ for the monofacial BSR cell) and assuming an internal electroluminescent efficiency of 10% (panel a) and 100 % (panel b). Figure 3 shows the efficiency (panels a and b) and the power density (panels d and e) for both bifacial and BSR TPV cells as a function of OBA and internal electroluminescent efficiency. Panels c and d in this figure show the increase on the bifacial cell maximum temperature (panel c) and the dissipated heat from its edges (panel f) as a function of the same parameters. It worth noting that the temperature of the bifacial cell is maximum at its center and its value depends on the emitter temperature, the optical and electric losses in the cell, and the cell dimensions (thickness and width). Panels c and f in Figure 2 show how TPV cell peak temperature and dissipated heat increase at high optical and electrical losses, as expected. A more detailed discussion on the temperature of the bifacial TPV cell can be found in the **Appendix**. Calculations in both figures assume a black body emitter at 1000ºC, a 200 μm thick and 1 cm wide bifacial TPV cell, and BSR cells operating at the same average temperature than bifacial cells.

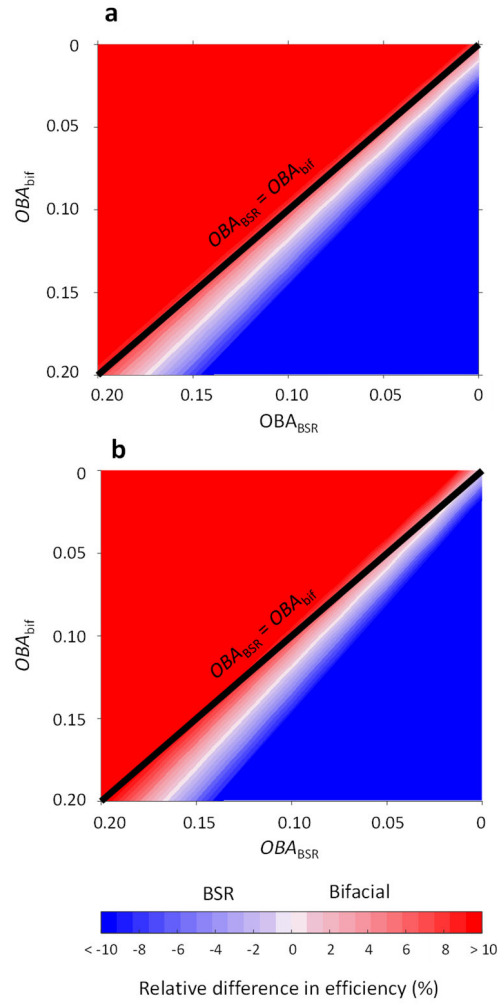

Figure 2. Relative difference in TPV efficiency between bifacial and BSR TPV cells as a function of the outband absorptance (OBA) of each kind of cell. OBA of BSR cells is equal to $1 - \rho_{BSR}$. Panel a and b assume an internal electroluminescence efficiency of 10 % and 100%, respectively. Red colored regions represent situations where bifacial TPV cells provide higher efficiency than BSR ones. The solid black line in both panels represent the case $OBA_{BSR} = OBA_{bif}$ (same amount of outband losses in both BSR and bifacial cells). Calculations assume a black body emitter at 1000ºC, a 200 μm thick and 1 cm wide bifacial TPV cell, and the same temperature for both bifacial and BSR cells.

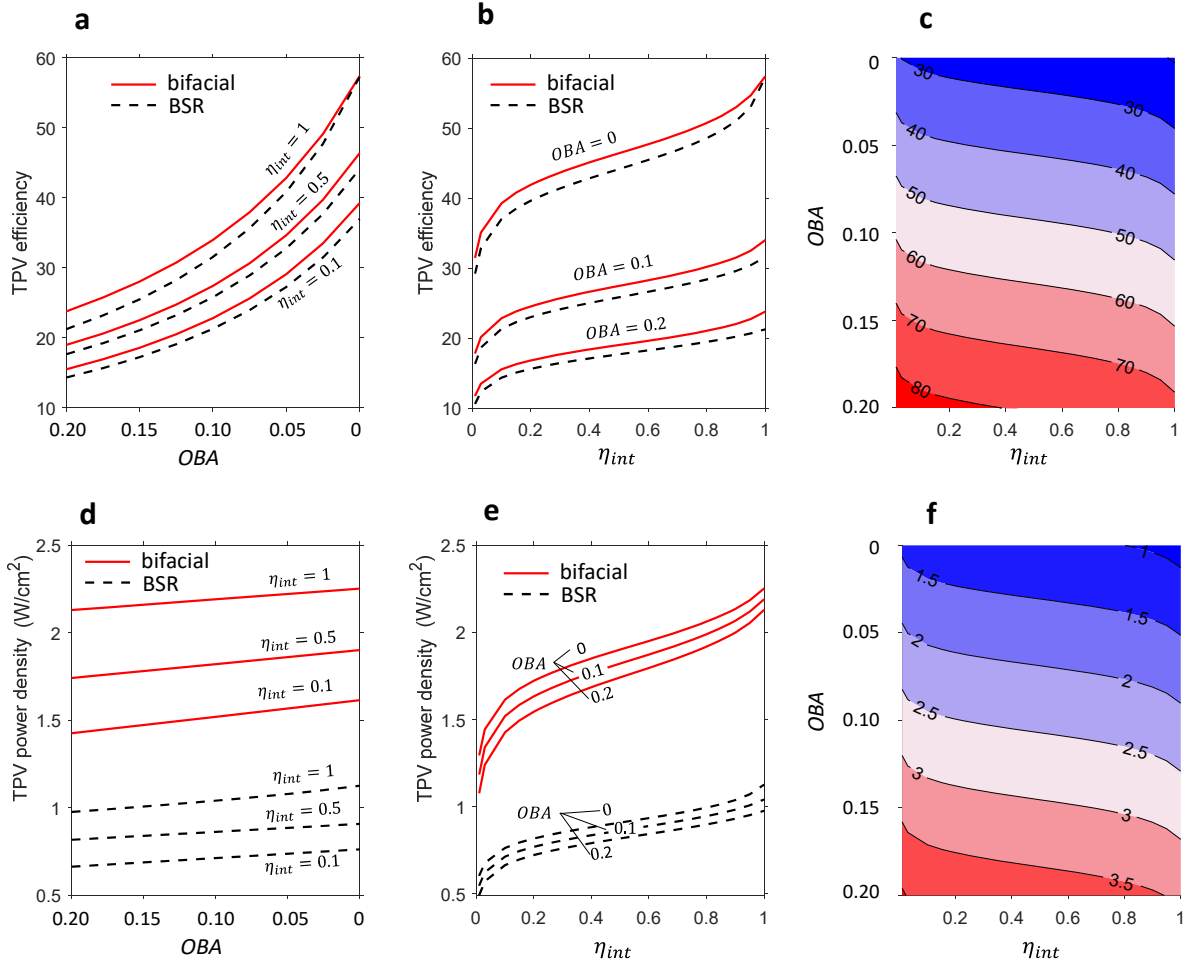

Figure 3. TPV efficiency (panels a and b) and power density (panels d, e) of bifacial and BSR cells as a function of the outband absorptance (OBA) and the internal electroluminescent efficiency ($\eta_{int}$). Panel c shows the maximum temperature difference at the center of the bifacial TPV cell with respect to the edge temperature, and panel f shows the required heat dissipation at the edges of the bifacial TPV cell, in W/cm. Calculations assume a black body emitter at 1000ºC, a 200 μm thick and 1 cm wide bifacial TPV cell, and both kinds of cells operating at the same average temperature of the bifacial TPV cell.

The results in Figure 2 show that bifacial TPV cells can provide higher efficiencies than BSR cells having similar amounts of OBA losses. Equivalently, bifacial cells provide the same efficiency than BSR cells even with higher amount of OBA losses. Only when both the mirror and the semiconductor are extremely efficient ($OBA_{BSR} \to 0$ and $\eta_{int} \to 1$) BSR cell approaches the conversion efficiency of the bifacial one (top-right corner in panel b of Figure 2, or top-right corners in panels a and b in Figure 3). This means that bifacial cells are less sensitive to optical and non-radiative recombination losses. As an example, a BSR TPV cell made of a semiconductor with 10% internal electroluminescence efficiency and with 5 % OBA ($\rho_{BSR}$ =0.95) would provide the same efficiency than a bifacial cell made of the same semiconductor and having a higher OBA of 6.4%. If the bifacial cell had the same OBA than the BSR one (i.e., 5%) it would provide an efficiency ~ 7 % (relative) higher. If the internal electroluminescent efficiency were 100%, the bifacial cell would require an OBA of 6 % to reach the same efficiency than a BSR cell with OBA of 5 %, and the bifacial efficiency when both cells have the same OBA (of 5%) would be ~ 5 % (relative) higher.

The reasons behind the higher efficiency of bifacial cells are 1) the higher output voltage that results from the higher photocurrent generation, and 2) the better photon recycling within the semiconductor. The first reason is linked to the higher generation/recombination ratio that is possible with bifacial TPV cells, i.e., $K_{bi} = 2$ in equation 5 that makes the negative term related to non-radiative recombination less noticeable; thus, enabling higher output voltages. The second reason is linked to the fact that electroluminescent photons are emitted within the semiconductor with a very high brightness (n²). Therefore, even a very small absorption in the mirror/semiconductor interface ($\rho_{BSR}$ <1) will significantly deteriorate the photogenerated current at a certain operating voltage. In a bifacial cell, most of the electroluminescent emitted photons are recycled, and only 1/n² of them leave the semiconductor from each side (i.e., $K_{mo} = 0$ in equation 5). Therefore, a bifacial cell is equivalent to a monofacial cell with an ideal mirror in terms of electroluminescence photon recycling. Indeed, only if both the mirror and the semiconductor are ideal ($OBA_{BSR} = 1 - \rho_{BSR} \to 0$ and $\eta_{int} \to 1$) both BSR and bifacial cells result in the same output voltages that maximize the generated power density (see equation 5), and therefore, only in this case they provide the same conversion efficiencies (as observed in panel b of Figure 2 and panels a and b of Figure 3). It is worth recalling that the output voltage is calculated to maximize the power density. When $\eta_{int} = \rho_{BSR} = 1$ in equation 5, only the first term remains for both bifacial and monofacial cells, and the voltage that maximizes the power density does not depend on the value of $K_{bi}$. This is why both bifacial and monofacial cell provide identical optimum voltages in this case, despite bifacial cells produce twice the power density.

Another obvious advantage of bifacial cells is that they provide a higher power density (panels d and e in Figure 3). In fact, the gain in power density is greater than 2 due to the increased voltage operation as explained above. Only when $OBA_{BSR} \to 0$ and $\eta_{int} \to 1$ the bifacial cell produces exactly twice the power density of the BSR one. Otherwise, bifacial cells can produce up to 2.2 times the power of the BSR cell (for the range of parameters explored in this work). This gain is more noticeable at low internal electroluminescent efficiencies and high OBA losses, as the effect of enhanced photogeneration is more significant in this case.

The discussion above focused on the case that the BSR cell is operated at the same temperature than the bifacial one. However, BSR-TPV cells enable a more efficient cooling from the rear side, and thus, they could be kept at a colder temperature (27ºC). In this case, BSR-TPV cells may provide higher efficiencies. Cooling issues of bifacial TPV cells are more significant when losses are high (high OBA and low $\eta_{int}$) and when the emitter temperature is also high. However, cooling can be improved by reducing the width and increasing the thickness of the cell, so that the high lateral heat flux within the cell do not result in an excessively high temperature (see **Appendix**).

Figure 4 shows the relative difference in TPV efficiency between BSR cells kept at 27ºC and bifacial cells having different widths (*w* = 1 cm for panels b and c, and *w* = 0.5 cm for panels d and e) as a function of the emitter temperature and the internal electroluminescence efficiency (panels b and d) and the outband absorption losses (panels c and e). Panel a shows the temperature profile in the cell for the six cases indicated with starts in the panels b-e. The cases A, B and C correspond to a 1-cm wide bifacial cell, whereas the cases D, E and F correspond to a 0.5-cm wide bifacial TPV cell. Results in Figure 4 illustrate that BSR cells tend to be more efficient at high emitter temperatures due to the better cooling. However, bifacial TPV cells can still be more efficient at high emitter temperatures by reducing the cell width (panels d and e), so that the temperature in the cell is kept low (panel a, cases F, E and D).

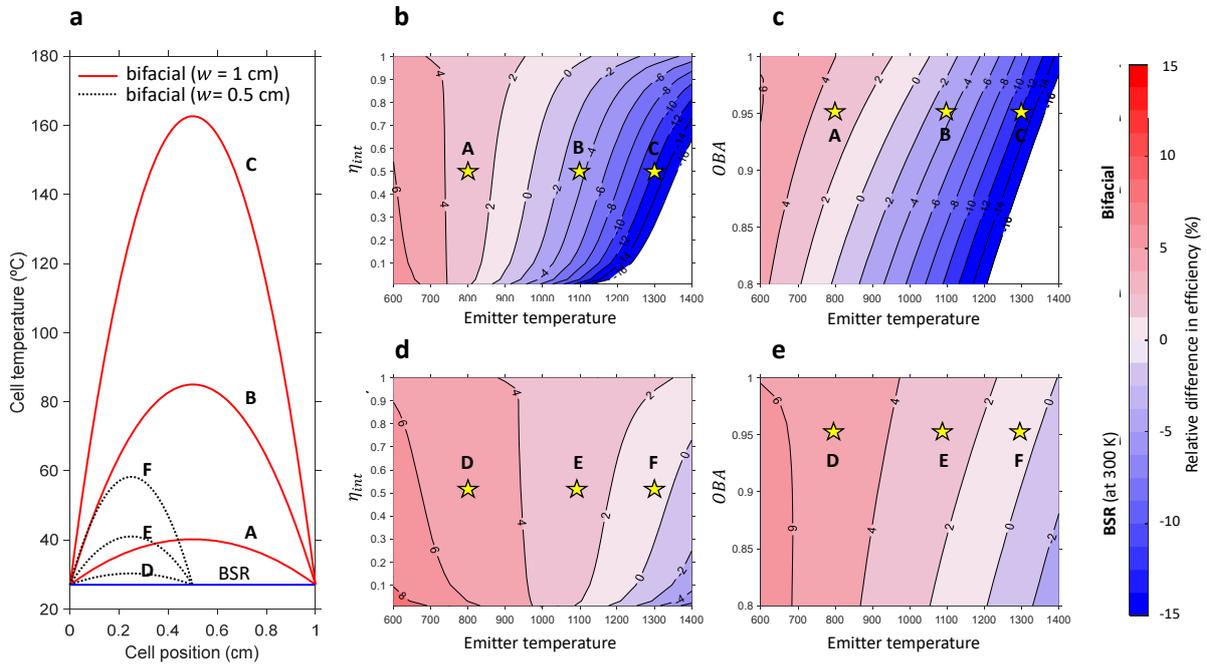

Figure 4. Panel a shows the temperature profile in BSR (solid blue) and bifacial TPV cells having widths of 1 cm (solid red) and 0.5 cm (dashed black) for the six cases indicated with stars in panels b to e. Panels b to e show the relative difference in TPV efficiency between bifacial and BSR TPV cells as a function of the emitter temperature and the internal electroluminescence efficiency (panels b, d) or the outband absorption losses (panels c, e). Panels b and c show the case of bifacial TPV cells having width of 1 cm, whereas panels d and f show the case of bifacial TPV cells having a width of 0.5 cm. Panels b and d assume an outband absorption loss of 5%, whereas panels c and e assume an internal electroluminescent efficiency of 50%.

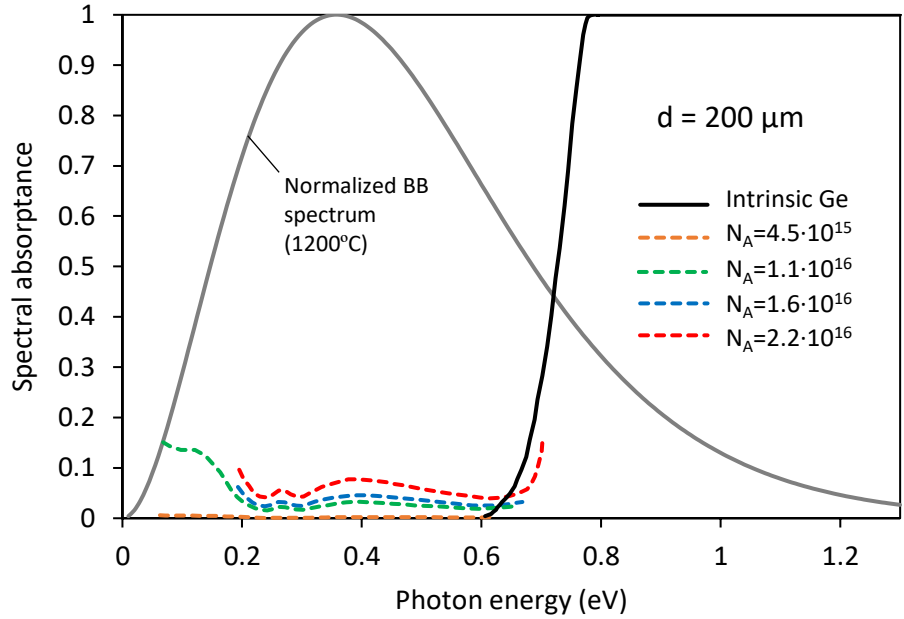

Figure 5. Outband spectral absorptance, i.e., $OBA(\lambda)$, of 200 μm thick p-Ge as a function of the doping level. The normalized black-body (BB) spectrum at 1200ºC is superimposed for reference. The values are calculated from the absorption coefficients of p-Ge reported in [17].

The discussion above implies that edge-cooled bifacial TPV cells must be designed for a maximum emitter temperature operation considering all kind of losses (optical and electrical). The key design parameters are the thickness and the width of the cell. Thicker cells can accommodate higher lateral heat fluxes within the cell, but outband optical absorption would increase. Thus, an optimal thickness will emerge for each kind of semiconductor material that minimize OBA while keeping a reasonable cell temperature.

As an example, Figure 5 shows the spectral outband absorptance, i.e., $OBA(\lambda)$, for 200 μm thick p-Ge, as calculated from the absorption coefficients reported in at different doping levels [17], [18]. P-Ge is selected because it is the typical substrate of choice for the manufacturing of Ge-based TPV cells [19], [20]. The overall OBA (for photon energies less than 0.66 eV) ranges from ~ 0.2 % for lightly doped p-Ge ($4.5 \cdot 10^{15}$ cm$^{-3}$) and high emitter temperatures (1400ºC) to ~ 6 % for more heavily doped p-Ge ($2.2 \cdot 10^{16}$ cm$^{-3}$) and lower emitter temperatures (800ºC).

Lower emitter temperatures are penalized by the higher overlap between the black body spectrum and the far-IR absorption of the Ge substrate. Higher doping is penalized by a higher free-carrier absorption in Ge. OBA could be reduced to 0.1 – 3% range by halving the thickness of the Ge cell (from 200 to 100 μm). However, narrower cells would be needed to keep a relatively low cell temperature, as explained above (see **Appendix**).

As observed in Figure 5, far-IR free-carrier absorption in Ge is significant and could become a serious limitation for heavily doped substrates or when the cell operates in high injection regime due to a very intense illumination [17], [21]. An alternative is to use thinner TPV cells made of direct-bandgap semiconductors (e.g., InGaAs [6]) that are bonded to a highly transparent substrate (e.g., Si or InP) that is electrically conductive or has lateral conduction layers, similar to those used in monolithic interconnected modules [22]. This design would also benefit from the higher internal electroluminescence of direct semiconductors. Nonetheless, the optimization of a specific bifacial cell design is out of the scope of this work and will be assessed in future investigations.

**Conclusions**

An edge-cooled bifacial TPV cell capable of producing about twice the power density of conventional BSR-TPV cells has been established. The enhanced photogeneration and improved photon recycling also enables higher output voltages and conversion efficiencies and makes them less sensitive to optical and electrical losses. By cooling the cells from the edges, it is possible to keep them at a relatively low temperature, provided that the cell size (width and thickness) is properly designed to accommodate the high lateral heat fluxes that are expected, especially when the emitter temperature is high. Another key advantage of the bifacial cell design is that it enables the development of relatively simple and highly packed TPV module designs using fewer materials, which may enable lower cost and higher efficiencies at module and system levels.

## Acknowledgements

This work has been funded by the project GETPV (PID2020-115719RB-C22) funded by the Spanish Ministry of Science and Innovation; and by the project MADRID-PV2 (CAM S2018/EMT-4308) funded by the Regional Government of Madrid.

**Appendix: Heat dissipation in edge-cooled bifacial TPV cells**

Panel d in Figure 6 shows the temperature profile in 0.5 and 1 cm wide cells as a function of the cell thickness ($d$ = 100 to 200 μm) for a fixed emitter temperature (1000ºC) and assuming no outband absorptance and 100% internal electroluminescence efficiency. Peak temperatures lower than 65ºC are found in all the cases at the center of the cell in this case. This peak value decreases with the cell thickness and increases with the cell width, as it could be expected. Therefore, a trade-off must be established between these two parameters to avoid excessive cell temperatures: thin cells are possible if they are narrow or, alternatively, wide cells are possible if they are thick. This is further illustrated in panels a, b, and c in Figure 7, which shows the difference between the peak and edge temperatures as a function of the cell width and thickness for three emitter temperatures: 800ºC (panel a), 1000ºC (panel b) and 1200ºC (panel c). The range of tolerable values for cell thickness and width narrows at high emitter temperatures, where thicker, narrower cells are needed. Panels d, e, and f in Figure 7 show the heat dissipation needed at the edges of the cell (i.e., $wQ_{tot}/2$), in W/cm. The heat dissipated increases with the width of the cell and, to a lesser extent, with its thickness. The required heat dissipation ranges from few hundreds of mW/cm (at 800ºC emitter temperature) to few W/cm (at 1200ºC emitter temperature). Assuming a typical heat transfer coefficient for cooling tubes of 0,1 W/cm²K this range of heat dissipation is attainable with cooling tubes of few mm diameter.

The effect of outband optical absorption and internal non-radiative recombination on the cell temperature and heat dissipation is shown in panels c and f of Figure 3. As expected, both the cell temperature and the heat dissipation increase with both kinds of losses. Thus, the final cell size (width and thickness) must be decided considering all these parameters.

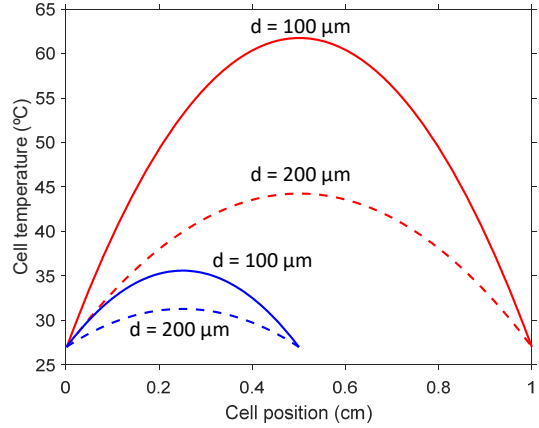

Figure 6. Calculated temperature profile in four kinds of bifacial TPV cells (0.5 and 1 cm wide, 100 and 200 μm thick) when irradiated by a 1000ºC black body emitter, assuming non outband optical losses ($OBA_{bif} = 0$), 100% internal electroluminescence efficiency, and a perimeter temperature (at x = 0 and x = w) of 27ºC.

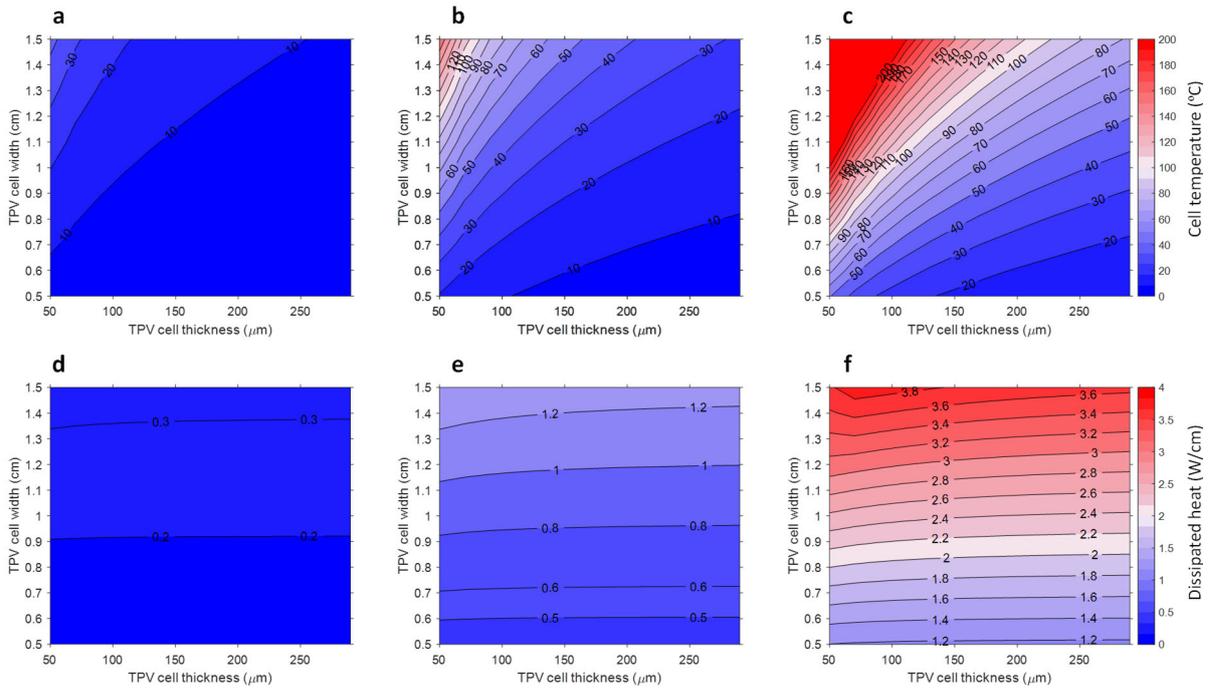

Figure 7. Maximum temperature increment at the center of the bifacial TPV cell in ºC (a, b, c) and the corresponding dissipated heat at the edges in W/cm (d, e, f) as a function of the cell thickness and width for emitter temperatures of 800ºC (a, d), 1000ºC (b, e) and 1200ºC (c, f). Calculations assume 100% outband transmittance ($OBA_{bif} = 0$), 100% internal electroluminescence efficiency, and a cell edge temperature of 27ºC.